\documentclass[prl,twocolumn,showpacs,preprintnumbers,amsmath,amssymb,floats]{revtex4-1}
\usepackage{graphicx}
\usepackage{amsmath}
\usepackage{epstopdf}
\usepackage{amsbsy}
\usepackage{color}
\usepackage{dcolumn}
\usepackage{bm}

\def\vk{{\bf k}}

\begin{document}

\title{Emergent defect states as a source of resistivity anisotropy in the nematic phase of iron pnictides}
\author{Maria N. Gastiasoro$^1$, I. Paul$^2$, Y. Wang$^3$, P. J. Hirschfeld$^3$, and Brian M. Andersen$^1$}
\affiliation{$^1$Niels Bohr Institute, University of Copenhagen, Universitetsparken 5, DK-2100 Copenhagen,
Denmark\\
$^2$Laboratoire Mat\'{e}riaux et Ph\'{e}nom\`enes Quantiques (UMR 7162 CNRS),
Universit\'{e} Paris Diderot-Paris 7, Bat. Condorcet, 75205 Paris Cedex 13, France\\
$^3$Department of Physics, University of Florida, Gainesville, Florida 32611, USA}

\date{\today}

\begin{abstract}
We consider the role of potential scatterers in the nematic phase of Fe-based superconductors
 above the transition temperature to the $(\pi,0)$ magnetic state but below the orthorhombic
structural transition.   The anisotropic spin fluctuations in this region can be frozen by disorder,
to create elongated magnetic droplets whose anisotropy grows as the magnetic transition is
approached.  Such states act as strong anisotropic  defect potentials which scatter with much higher probability
perpendicular to their length than parallel, although the actual crystal symmetry breaking is tiny.  We
calculate the scattering potentials, relaxation rates, and conductivity in this region, and show that
such emergent defect states are essential for the transport anisotropy observed in experiments.
\end{abstract}


\maketitle

The  origin of electronic nematic behavior, i.e.  spontaneous breaking of discrete rotational symmetry  preserving translational symmetry, is one of the most fascinating questions in the field of Fe-based superconductivity,
involving the interplay of magnetic, orbital, and ionic fluctuations.   Strong in-plane anisotropy has been reported  in transport\cite{tanatar,chu,ying,chu2,blomberg,ishida13,kuo14}, angular resolved photoemission (ARPES)\cite{yi11}, neutron scattering\cite{zhao}, optical spectroscopy\cite{nakajima,dusza}, Andreev point contact\cite{LGreene} and torque magnetometry\cite{kasahara} measurements.   Since the various fluctuation channels in these multiband systems  all couple to one another below the tetragonal to orthorhombic structural transition that occurs at $T_s$ in many systems, all response functions become anisotropic and  it is not easy to decide which fluctuations drive the ordering nematic phenomena observed. Theoretically, both  spin nematic and  orbital scenarios   have been proposed\cite{Fernandes14}.

In systems with large spin nematic susceptibility, strong anisotropy is expected in the spin fluctuations in the orthorhombic phase below $T_s$, even if the structural anisotropy  is small.
Such anisotropy will certainly influence transport properties; this is the basis of theories of transport by several groups\cite{Schmalian11,Prelovsek14,Brydon14}, arguing that at $T_s$, the magnetic correlation length becomes anisotropic and drives the anisotropy in the electronic inelastic scattering rate. Disorder is described entirely through a momentum-independent scattering rate and is required only to short-circuit ``cold spots" on the Fermi surface.

In the present work, the transport anisotropy of the nematic phase is also explained via spin fluctuation anisotropy, but {\it through the generation of strongly anisotropic impurity states.}  Our work is motivated by the observation by many STM experiments
of  $C_4$ symmetry breaking around point defects locally\cite{chuang10,song11,zhou11,grothe12,hanaguri12,allan13,Rosenthal14}; these experiments can  exhibit effects that are missed by average bulk probes.   In fact,  in some systems evidence for nematic symmetry breaking in the form of highly anisotropic $C_2$ defect states  is seen  in the nominally tetragonal phase above $T_s$\cite{Rosenthal14}.  These responses are generally attributed to residual local strains which break $C_4$ symmetry locally, together with a large residual nematic susceptibility.\cite{PJH_Davis} In the ordered stripe ($\pi,0$)
magnetic phase below the N\'eel temperature $T_N$ in many of the parent and underdoped materials, the $C_4$ symmetry
is broken by the magnetism itself.  Nevertheless the symmetry breaking of the electronic structure around local Co defects in lightly doped Ca122 was observed to be so enhanced that this result was cited as the first evidence for a strong nematic tendency in these systems\cite{chuang10}.  In addition, it was suggested in Ref.~\onlinecite{allan13} that such ``nematogen"  defects could be responsible for the transport anisotropy.

Recently, we examined the microscopic origin of
 nematic  defect states in the ordered phase, and proposed that they result from the effect of a nonmagnetic impurity on the energy balance between two  magnetic phases, the ($\pi,0$) stripe ground state and a nearby ($\pi,\pi$) competing
N\'eel state\cite{Navarro_cigar}.    The relative stability of the latter at hole doping leads to an elongated  dimer-like structure in both charge distribution and low-energy local density of states (LDOS) in  agreement with experiments\cite{chuang10,allan13}.  $C_4$-broken impurity states were discussed earlier in the context of localized spin models\cite{chen09} and pinned fluctuating orbital order\cite{kontani12}, but in neither cases was the dimer-like structure seen in experiment reported.

The emergent nematogen defect states found in Ref. \onlinecite{Navarro_cigar} become $C_4$ symmetric above $T_N$ in the tetragonal phase. Yet transport anisotropy experiments in Ba122 exhibit significant anisotropy also in the ``nematic phase" $T_N<T<T_s$ where there is no long-range magnetic stripe order, or in the tetragonal phase in the presence of external stress.\cite{tanatar,chu,ying,chu2,blomberg,ishida13,kuo14} It is not clear, however, whether nematogens can form around point-like impurities in this phase, i.e. whether the anisotropic spin fluctuations in a spin-nematic scenario can condense around a defect to give a similar transport anisotropy in this case.

There are several key aspects of the transport experiments\cite{tanatar,chu,ying,chu2,blomberg,ishida13,kuo14} above $T_N$
that any theory needs to account for:  1) the counterintuitive sign of
the resistivity anisotropy on the electron-doped side, where
$\rho_b>\rho_a$ although $b<a$; 2) the possible sign change but also
significant decrease of the anisotropy on the hole-doped side\cite{blomberg}; 3) the
 decrease of the anisotropy upon annealing\cite{ishida13};  4) the pronounced
increase in $\rho_b$ as $T_N$ is approached, with little or no
increase in $\rho_a$; and 5) the decrease in anisotropy both with
increasing $T$ and electron overdoping.  We believe that
theories which propose transport anisotropy due to scattering of
electrons from spin fluctuations alone are able to account for only some
of these salient features, and that including the role of emergent
defect states in these correlated systems
provides a more natural explanation for the observations.

In this work, we discuss first the growth of anisotropic spin fluctuations in the nematic phase as $T_N$ is approached from above.    We extend the theory of impurity-induced emergent defects  states into the nematic phase with an unbiased microscopic calculation of the  local electronic structure near a point-like nonmagnetic  impurity potential in a situation
where the $C_4$ symmetry of the host bands has been broken very slightly below $T_s$. This gives rise to the {\it same} anisotropic spin fluctuations considered as the source of transport anisotropy by the authors of Refs.~\onlinecite{Schmalian11,Prelovsek14,Brydon14}, but impurities play a very different and essential role.
 We find that the impurity state in the nematic phase  is strongly anisotropic due to the enhanced background nematic response
  arising from electronic correlations.\cite{Andersen_nematic12}  Specifically, we calculate the momentum-dependent effective impurity potential,  scattering rate, and conductivity in the nematic phase.

The Hamiltonian is given by
\begin{equation}
 \label{eq:H}
\mathcal{H}=\mathcal{H}_{0}+\mathcal{H}_{oo}+\mathcal{H}_{int}+\mathcal{H}_{imp},
\end{equation}
where $\mathcal{H}_{0}$ denotes the kinetic energy
\begin{equation}
 \label{eq:H0}
\mathcal{H}_{0}=\sum_{\mathbf{ij},\mu\nu,\sigma}t_{\mathbf{ij}}^{\mu\nu}c_{\mathbf{i}\mu\sigma}^{\dagger}c_{\mathbf{j}\nu\sigma}-\mu_0\sum_{\mathbf{i}\mu\sigma}n_{\mathbf{i}\mu\sigma},
\end{equation}
with tight-binding parameters adopted from Ref.~\onlinecite{ikeda10}.
Here, ${\mathbf{i}}$ and ${\mathbf{j}}$ denote lattice sites, $\sigma$ the spin, and $\mu_0$ is the chemical potential that sets the doping level $x=0$.
The indices $\mu$ and $\nu$ are the five iron $d$ orbitals.
$\mathcal{H}_{oo}=\frac{\delta}{2}\sum_{\mathbf{i}}\left( n_{\mathbf{i}yz}-n_{\mathbf{i}xz}\right)$ mimics the orthorhombicity of the band below $T_s$, for a non-zero $\delta$ orbital order parameter. We have also studied $C_2$ symmetric bands arising from hopping anisotropy and found similar results to those reported below.
The third term in Eq.(\ref{eq:H}) describes the Hubbard-Hund interaction
\begin{align}
 \label{eq:Hint}
 \mathcal{H}_{int}&=U\sum_{\mathbf{i},\mu}n_{\mathbf{i}\mu\uparrow}n_{\mathbf{i}\mu\downarrow}+(U'-\frac{J}{2})\sum_{\mathbf{i},\mu<\nu,\sigma\sigma'}n_{\mathbf{i}\mu\sigma}n_{\mathbf{i}\nu\sigma'}\\\nonumber
&\quad-2J\sum_{\mathbf{i},\mu<\nu}\vec{S}_{\mathbf{i}\mu}\cdot\vec{S}_{\mathbf{i}\nu}+J'\sum_{\mathbf{i},\mu<\nu,\sigma}c_{\mathbf{i}\mu\sigma}^{\dagger}c_{\mathbf{i}\mu\bar{\sigma}}^{\dagger}c_{\mathbf{i}\nu\bar{\sigma}}c_{\mathbf{i}\nu\sigma},
\end{align}
including  the intraorbital (interorbital) on-site repulsion $U$ ($U'$), the Hund's coupling $J$ and the pair hopping energy $J'$. We assume $U'=U-2J$ and $J'=J$ and fix $U=1.0$ eV and $J=U/4$.
Finally, $\mathcal{H}_{imp}=V_{imp}\sum_{\mu\sigma}c_{\mathbf{i^*}\mu\sigma}^{\dagger}c_{\mathbf{i^*}\mu\sigma}$ is the impurity potential, adding a potential $V_{imp}=1.5$eV at the impurity site $\mathbf{i^*}$. We neglect the orbital dependence of the impurity potential for simplicity.
After mean-field decoupling of Eq.~\eqref{eq:Hint}, we solve the following eigenvalue problem $\sum_{\mathbf{j}\nu}
H^{\mu\nu}_{\mathbf{i} \mathbf{j} \sigma}
 u_{\mathbf{j}\nu\sigma}^{n}
=E_{n\sigma} u_{\mathbf{i}\mu\sigma}^{n}$,
where
\begin{align}
&H^{\mu\nu}_{\mathbf{i} \mathbf{j} \sigma}=t_{\mathbf{ij}}^{\mu\nu}+\delta_{\mathbf{ij}}\delta_{\mu\nu}[-\mu_0+\delta(\delta_{\mu yz}-\delta_{\mu xz})+\delta_{\mathbf{ii^*}}V_{imp}\nonumber\\
&+U \langle n_{\mathbf{i}\mu\bar{\sigma}}\rangle+\sum_{\mu' \neq \mu}(U'\langle n_{\mathbf{i}\mu' \bar{\sigma}}\rangle+(U'-J)\langle n_{\mathbf{i}\mu' \sigma}\rangle)],
 \end{align}
on a $30\times 30$ lattice with self-consistently obtained densities $\langle n_{\mathbf{i}\mu\sigma} \rangle=\sum_{n}|u_{\mathbf{i}\mu\sigma}^{n}|^{2}f(E_{n\sigma})$ for each site and orbital.

\begin{figure}[b]
\begin{center}
\includegraphics[width=8.0cm]{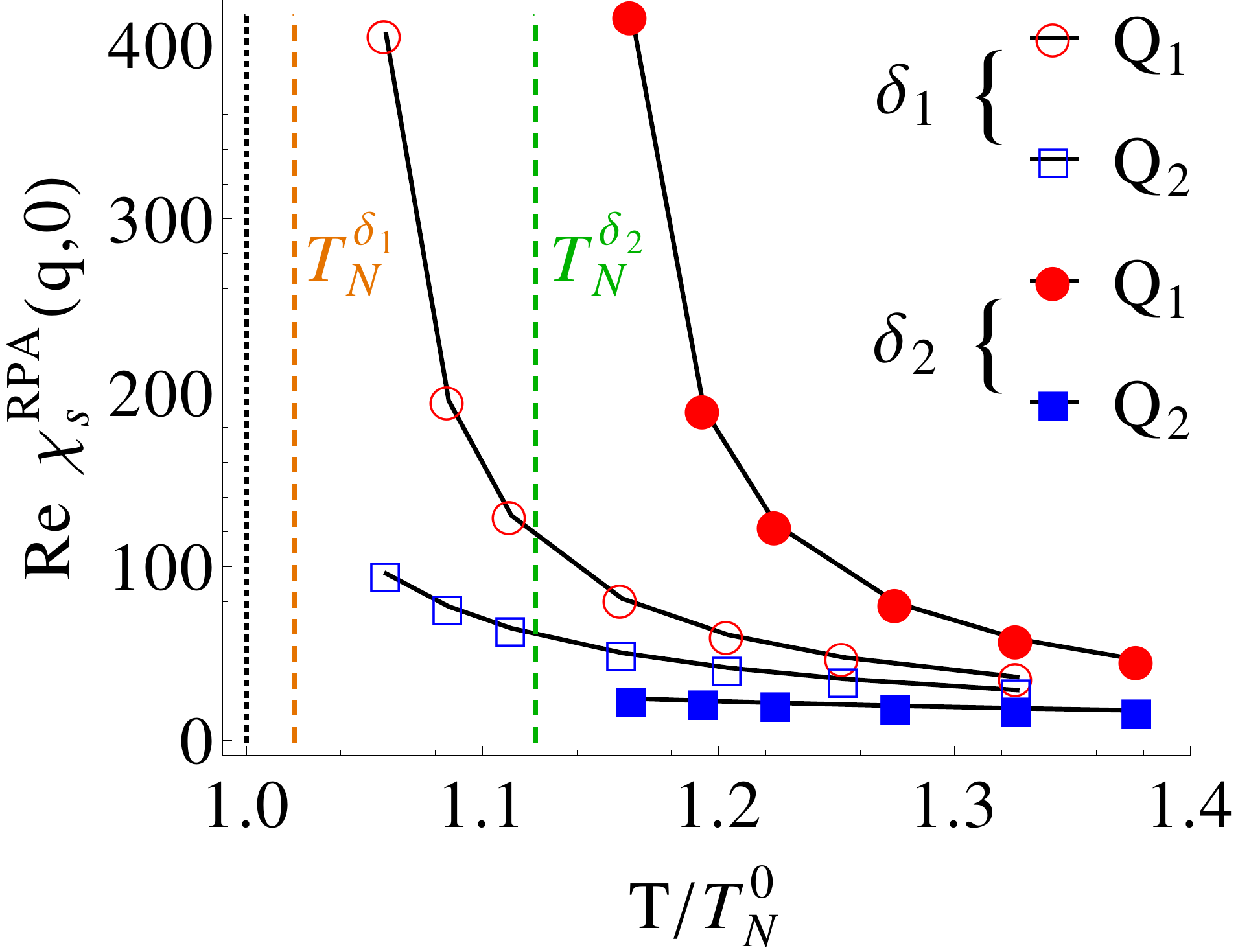}
\end{center}
\caption{(Color online) Real part of the homogeneous RPA spin susceptibility $\chi_s^{RPA}(\mathbf{q},0)$ at $\mathbf{Q_1}\equiv (\pi,0)$ (red curves) and $\mathbf{Q_2}\equiv (0,\pi)$ (blue curves) as a function of $T$ normalized to the N\'{e}el temperature $T_N^0$ of the tetragonal band ($\delta=0$). Open (solid) symbols refer to the degree of orbital order, $\delta_1=16$meV ($\delta_2=80$meV), and the dashed green (dashed orange) vertical lines indicate the corresponding relevant $T_N^\delta$.}
\label{fig:1}
\end{figure}

In the homogeneous orthorhombic ``nematic" phase above $T_N$, the important effect of the $xz$-$yz$ orbital splitting is to enhance (diminish) the spin susceptibility at $\mathbf{Q_1}\equiv (\pi,0)$ ($\mathbf{Q_2}\equiv (0,\pi))$ as shown in Fig.~\ref{fig:1} for two cases with $\delta_1=16$ meV ($T_N^{\delta_1}$) and $\delta_2=80$ meV ($T_N^{\delta_2}$). The enhanced susceptibility at $\mathbf{Q_1}\equiv (\pi,0)$ pushes $T_N$ up. As seen explicitly from Fig.~\ref{fig:1}, even a small orbital splitting $\delta$ leads eventually to an arbitrarily large spin anisotropy upon approaching the instability.\cite{Andersen_nematic12}

\begin{figure}[]
\begin{center}
\includegraphics[width=8.0cm]{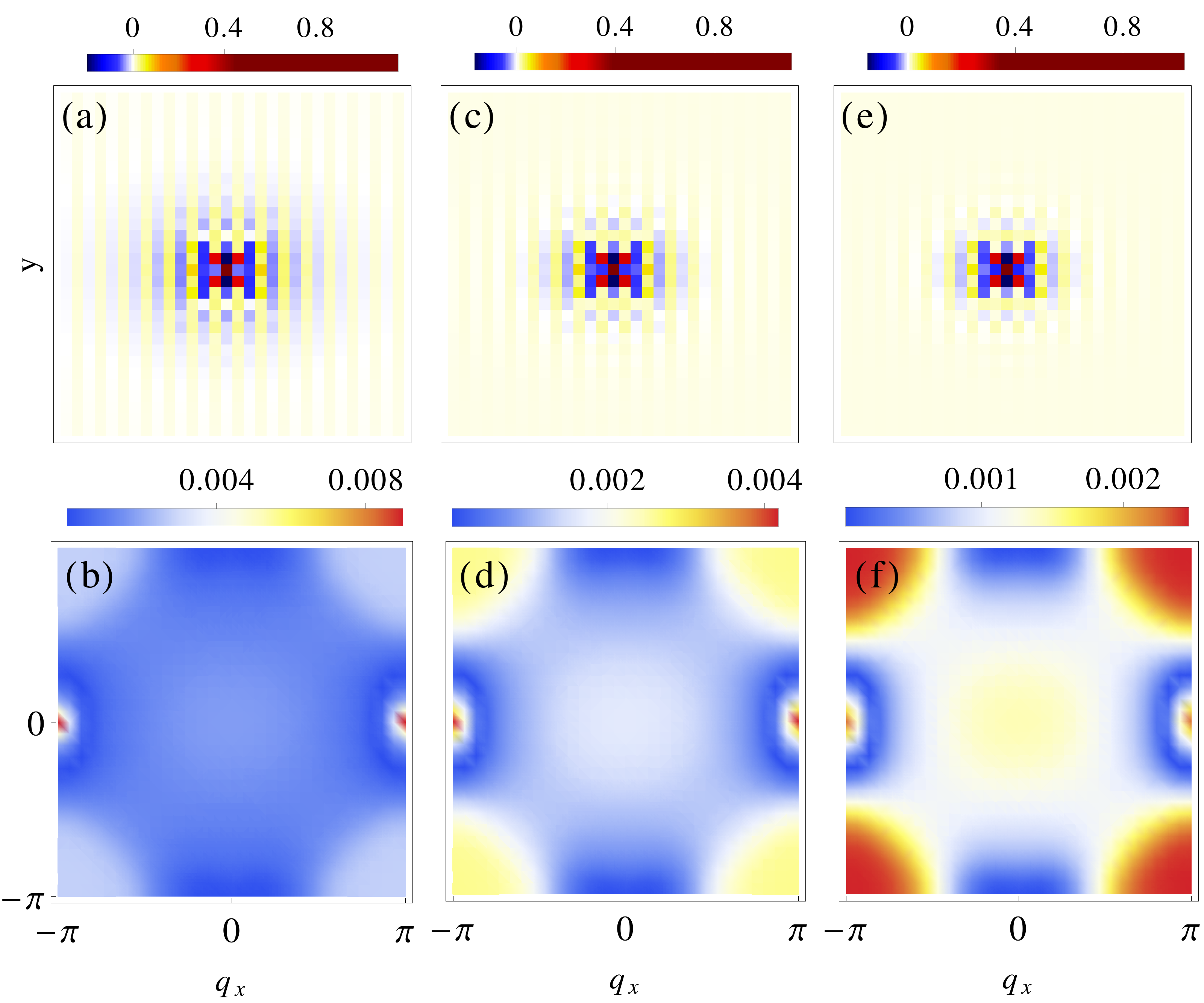}
\end{center}
\caption{(Color online) Real space magnetization $m(\mathbf{r})$ of a $V_{imp}=1.5$ eV impurity for $\delta_2=80$ meV at temperatures $T/T_N^{\delta_2}=$1.06 (a), 1.14 (c), and 1.23(e). (b), (d) and (f) show the Fourier transform $|m(\mathbf{q})|$ of (a), (c) and (e), respectively.}
\label{fig:2}
\end{figure}

How does the electronic structure near the impurities reflect the spin anisotropy of the nematic phase?
In Fig.~\ref{fig:2} we show local magnetization $m(\mathbf{r})$ nucleated by an impurity in the nematic state as a function of $T$. As seen, the emergent defect object pins  the order locally\cite{andersen07,Navarro_LiFeAs} and therefore incorporates the growing spin fluctuation anisotropy in the host upon approaching the magnetic instability. The growing $x$-$y$ anisotropy is clearly evident in the Fourier images in the lower row of Fig.~\ref{fig:2}. These impurity nematogens are the nematic phase equivalents of the nematogens studied in the SDW phase of Ref.~\onlinecite{Navarro_cigar}.

In order to determine the transport properties of the nematic defect states, we calculate first the scattering rate in the Born approximation
\begin{align}
\label{eq:tauinv}
 \frac{1}{\tau^l_{\mathbf{k}\alpha}}&=n_{imp}\frac{2\pi}{\hbar} \frac{1}{V} \sum_{\mathbf{k'}\beta} \left|\textrm{tr} \left( \hat\sigma_l \hat{\mathcal{V}}^{imp}_{\sigma\sigma'}(\mathbf{k}\alpha,\mathbf{k'}\beta) \right) \right|^2  \times \nonumber\\
 &\quad\delta(\epsilon_\mathbf{k\alpha}-\epsilon_\mathbf{k'\beta})\left( 1-\frac{\mathbf{v}_F^\alpha(\mathbf{k})\cdot\mathbf{v}_F^\beta(\mathbf{k'})}{|\mathbf{v}^\alpha_F(\mathbf{k})||\mathbf{v}_F^\beta(\mathbf{k'})|} \right),
 \end{align}
where $l=0$ ($l=3$) corresponds to the charge (magnetic) scattering rate and  $1/\tau_{\mathbf{k}\alpha}\equiv 1/\tau^0_{\vk \alpha}+1/\tau^3_{\vk \alpha}$ is the total scattering rate on band $\alpha$.
The term $\hat{\mathcal{V}}^{imp}_{\sigma\sigma'}(\mathbf{k}\alpha,\mathbf{k'}\beta) \equiv\langle \mathbf{k'}\beta\sigma'|\mathcal{V}^{imp}| \mathbf{k}\alpha\sigma\rangle\equiv \langle \mathbf{k'}\beta\sigma'|\mathcal{H}-\mathcal{H}_{(V_{imp}=0)}| \mathbf{k}\alpha\sigma\rangle$ is the matrix element of the impurity Hamiltonian for the fully converged self-consistent eigenvalue problem
\begin{equation}
 \hat{\mathcal{V}}^{imp}_{\sigma\sigma'}(\mathbf{k}\alpha,\mathbf{k'}\beta)=\sum_{\mu\nu} a_{\mathbf{k}\mu}^{\alpha*} \omega_{\mathbf{k}\sigma\mathbf{k'}\sigma'}^{\mu\nu} a_{\mathbf{k'}\nu}^\beta-\epsilon_{\mathbf{k}\alpha}\delta_{\mathbf{k}\mathbf{k'}}\delta_{\alpha\beta}.
\end{equation}
Here  $\omega_{\mathbf{k}\sigma\mathbf{k'}\sigma'}^{\mu\nu}=\frac{1}{N}\sum_{n} \sum_{\mathbf{i}\mathbf{j}}u_{\mathbf{j}\nu\sigma'}^{n*}  u_{\mathbf{i}\mu\sigma}^n E_{n \sigma} e^{-i\mathbf{k'}\mathbf{r_j}} e^{i\mathbf{k}\mathbf{r_i}} $,  and $a_{\mathbf{k}\mu}^\alpha$ are the matrix elements of the unitary  transformation from orbitals to bands.
Finally, ${\mathbf{v}}^\alpha_F(\mathbf{k})$ denotes the Fermi velocity of band $\alpha$, and the last term in parentheses in Eq.~(\ref{eq:tauinv}) is an approximation to the vertex corrections in the full Kubo formula  by Ziman\cite{Ziman} that has proven accurate for anisotropic scatterers\cite{LawrenceCole}.

\begin{figure}[b]
\begin{center}
\includegraphics[width=8.0cm]{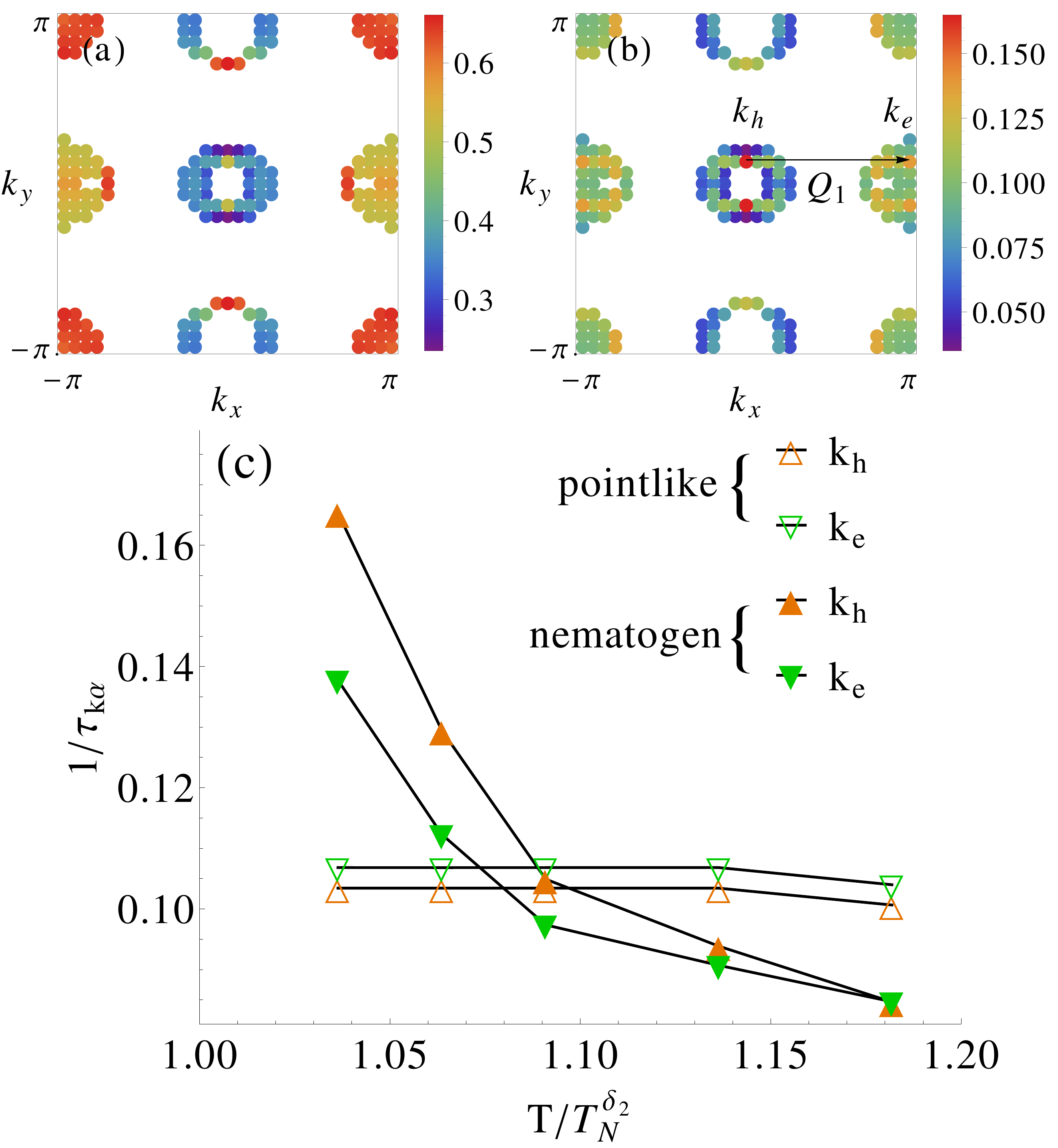}
\end{center}
\caption{(Color online) (a) Map of $1/\tau_{\vk\alpha}$ vs. $k_x,k_y$ for point-like $ \hat{\mathcal{V}}^{imp}_{\sigma\sigma'}(\mathbf{k}\alpha,\mathbf{k'}\beta)=\hat\sigma_0V_{imp}\sum_{\mu} a_{\mathbf{k}\mu}^{\alpha*} a_{\mathbf{k'}\mu}^\beta$ at $T/T_N^{\delta_2}=1.036$. Values are shown for all $\vk$ within a range $\sim 2 k_BT$ of the Fermi surface.
(b) Same map for nematogen with $\hat{\mathcal{V}}^{imp}_{\sigma\sigma'}(\mathbf{k}\alpha,\mathbf{k'}\beta)$ determined self-consistently. The arrow indicates the dominant $\mathbf{Q_1}$ scattering between the particle and hole pockets.
(c) Scattering rates from (a) [scaled by 1/5] and (b) at $\vk_{h}$, $\vk_{e}$ vs. $T$.
}
\label{fig:3}
\end{figure}

In Fig. \ref{fig:3}, we show the effect of local freezing of the spin fluctuations on the scattering rate anisotropy by plotting $1/\tau_{\vk\alpha}$ explicitly,
first for a point-like scatterer of potential $V_{imp}$ with no self-consistency in \ref{fig:3}(a). It is seen that the distribution of scattering weight reflects the
small orbital ordering that has created a slightly orthorhombic Fermi surface.   Since $V_{imp}$ is momentum independent, the
variation reflects primarily the band-orbital matrix elements for this model.    
Figure~\ref{fig:3}(b) now shows how the nematogen scattering rate reflects the intrinsic spin fluctuations in the system.  The 
localized object in real space couples fluctuations at all $\bf q$,  but these include important contributions from those scattering processes that dominate the
fluctuations in the homogeneous system, i.e. the scattering between like orbitals on hole and electron pockets as seen in \ref{fig:3}(b).
The point-like scatterer
leads to a scattering rate that is nearly $T$-independent, while the nematogen scattering rate grows as the magnetic transition is
approached, as shown in Fig. 3(c).  For the nematogen scattering, the
charge scattering rate  is also nearly $T$-independent. It is the magnetic scattering
rate that provides both the strong $T$-dependence and the enhanced anisotropy.

\begin{figure}[b]
\begin{center}
\includegraphics[width=8.0cm]{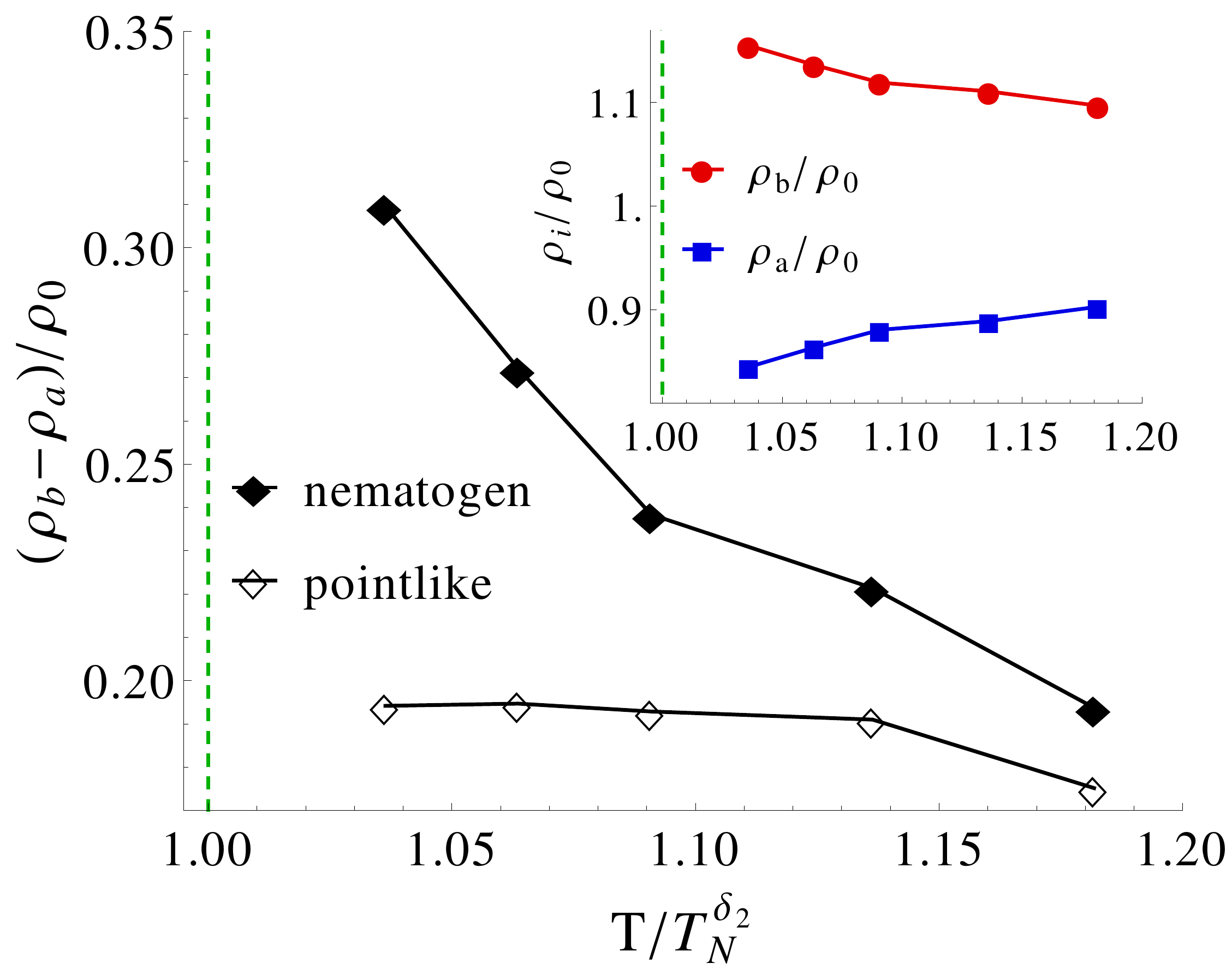}
\end{center}
\caption{(Color online) Resistivity anisotropy $\Delta \rho$ versus $T$ for non-selfconsistent point-like (empty diamonds) and selfconsistent nematogen (filled diamonds) impurity scatterers.
Inset shows the $T$-dependence of $\rho_a/\rho_0$ (circles) and $\rho_b/\rho_0$ (squares) for the nematogen case. }
\label{fig:4}
\end{figure}

Turning finally to the conductivity obtained from
\begin{equation}
  \mathbf{\sigma}_{ij}=e^2 \frac{1}{V}\sum_{\mathbf{k}\alpha} \mathbf{v}_i^\alpha(\mathbf{k}) \mathbf{v}_j^\alpha(\mathbf{k}) \tau(\epsilon_{\mathbf{k}\alpha})\left( -\frac{\partial f}{\partial \epsilon_{\mathbf{k}\alpha}} \right),
  \end{equation}
we show in Fig.~\ref{fig:4} the resistivity anisotropy $\Delta\rho=(\rho_b-\rho_a)/\rho_0$ as a function of $T$ with $\rho_0=(\rho_a+\rho_b)/2$. As expected from Fig.~\ref{fig:3}, the anisotropy in the case of point-like scatterers is essentially $T$-independent and caused only by the band. On the other hand, for the nematogens  $\Delta \rho$ rises rapidly upon approaching the magnetic instability, in agreement with experiments.  As $T_N$ is approached, the divergence of the spin fluctuation scattering rate is cut off eventually:
in our simulation by the system size, in the real sample by the inter-nematogen distance.

With the above results in hand, we can explain the key properties 1)-5) of the transport
experiments discussed in the introduction. Our picture assumes that the Ba122 system, in
particular, contains a great deal of disorder away from the FeAs plane which determines the large
values of the resistivity near $T_N$.   This is consistent with the large constant $\rho(T_N)$ and small $T^2$ coefficient in the parent and lightly doped
materials. These scatterers are
weak, do not pin low-energy spin fluctuations, and cannot contribute to the resistivity anisotropy.  In
the parent compound even after annealing, a few vacancies in the FeAs plane creating stronger scatterers
remain, and give rise to a small peak in the $b$-axis resistivity above $T_N$ due to nematogen
formation.   Upon doping with Co, however, the concentration of nematogens rises quasi-linearly  and enhances
the resistivity anisotropy and peaks in $\rho_b$, as seen in experiment, until the critical
doping where $T_N$ goes to zero and the magnetic fluctuations which drive the anisotropy weaken. Hole doping with K, on the
other hand, introduces much weaker out-of-plane scatterers that cannot induce nematogens;\cite{Navarro_cigar} the
anisotropy is then essentially zero, with the exception of that driven by few residual vacancies.
We have checked that within our model the sign of the anisotropy indeed 
changes on the hole doped side as in experiment, but this is a band-structure dependent effect;  the more important
effect, in our view, is the dramatic collapse of the anisotropy also observed in the hole doped system\cite{blomberg}.

We emphasize again that the physics of resistivity anisotropy in our view arises ultimately from the
same anisotropy in the spin fluctuation spectrum  invoked by the authors of Refs.~\onlinecite{Schmalian11,Prelovsek14,Brydon14}. Nevertheless, the
importance of these fluctuations in the current picture is that they condense into an emergent
defect state above $T_N$ whose anisotropy grows in response to the small orthorhombic symmetry
breaking below $T_s$, which then scatters electrons anisotropically.  We have shown that a tiny Fermi
surface asymmetry, reflected in a very weak anisotropy of the Drude weight\cite{footnote}, is dramatically enhanced
by spin fluctuations near $T_N$ such that scattering rate anisotropies of order 100\% are possible.

Strong evidence in favor of this picture comes from the annealing experiments of Ishida {\it et al.}\cite{ishida13},
who show that when strong disorder is removed the anisotropy drops, and attribute the remaining anisotropy to Co
atoms, as we do here.   While a reduction in anisotropy with decreasing disorder  is also possible in
the inelastic scattering models, as pointed out e.g. by Breitkreitz {\it et al.}\cite{Brydon14}, it occurs in a parameter regime
where spin fluctuation scattering dominates elastic scattering, in contrast to the situation in
experiments. 

While these theories seem to
account for the dramatic reduction of
anisotropy on the hole-doped side, this agreement depends on the
ellipticity of the 2D electron bands assumed.  In the Ba122 system,
however, the electron pockets have an ellipticity that changes sign with $k_z$, leading to a near-cancellation
 of band structure contributions to the anisotropy. The scattering rate anisotropy due to the nematogens, on 
 the other hand, depends uniquely on the orthorhombicity, rather than special features of the band. We note that the nematic susceptibility measured in this material is quite
electron-hole symmetric.\cite{Boehmer}
 
Recently, Kuo and Fisher\cite{kuo14} criticized the idea of an extrinsic source of the 
anisotropy, since samples with very different RRRs have similar resistivity anisotropies, and 
different chemical substituents corresponding to the same doping exhibit similar anisotropies as well.
Neither of these observations contradicts our analysis, however, since first, the large differences in sample
quality and RRR are caused largely by out-of-plane disorder that does not create nematogens.  Second, for potentials
strong enough to create nematogens,  the anisotropy in the scattering rate arises from the spin fluctuations themselves;
the strength of the potential for different impurities affects mainly the magnitude of the average resistivity and much less its anisotropy.

In summary, we have discussed an impurity-driven scenario for the remarkable transport anisotropy
in Fe-based superconductors that explains all essential features of these measurements, and argues for
an increased focus on the unusual role played by impurities in these systems with strong spin fluctuations
near a magnetic transition.

We thank R. Fernandes and S. Mukherjee for useful discussions. B.M.A. and M.N.G. acknowledge support from Lundbeckfond fellowship (grant A9318). P.J.H. and Y.W. were supported by NSF-DMR-1005625.


\begin{thebibliography}{00}
\bibitem{footnote} The ratio $(\langle v_x^2\rangle-\langle v_y^2\rangle)/\langle v_x^2\rangle$ is of order 10\% for the
parameters shown in Fig. \ref{fig:3}.
\end{thebibliography}
\end{document}